\documentclass[10pt]{iopart}
\usepackage{graphicx,epsfig}
\usepackage{bm}
\usepackage{iopams}
\usepackage{color,ulem}
\newcommand\eqref[1]{(\ref{#1})}
\newcommand{\dd}{\textrm{d}}
\newcommand{\ba}{\begin{eqnarray}}
\newcommand{\ea}{\end{eqnarray}}

\usepackage{pdfsync}
\begin{document}

\title[Accelerated cosmic expansion without dark energy]{Cotton Gravity:  the cosmological constant as spatial curvature} 
\author{ Roberto A. Sussman$^\dagger$ and Sebasti\'an N\'ajera$^\dagger$}
\address{$^\dagger$ Instituto de Ciencias Nucleares, Universidad Nacional Aut\'onoma de M\'exico (ICN-UNAM),
A. P. 70--543, 04510 M\'exico D. F., M\'exico.}
\eads{$^\ddagger$\mailto{sussman@nucleares.unam.mx}}
\date{\today}
\begin{abstract} 
We derive Friedman-Lema\^\i tre-Robertson-Walker (FLRW) models as non-trivial solutions of ``Cotton Gravity" (CG), a recently proposed gravity theory alternative to General Relativity (GR) based on the Cotton tensor.  Using  an equivalent formulation, we show that CG leads to FLRW models with a modified expression for spatial curvature in terms of the Ricci scalar of hypersurfaces orthonormal to the 4-velocity. Considering models compatible with a well posed initial value formulation leads to operationally the same FLRW models in GR, but endowed with a precise covariant characterization of the positive/negative cosmological constant as the case with constant negative/positive spatial curvature. Under CG, the $\Lambda$CDM model becomes the unique FLRW dust model with constant negative spatial curvature.                                                        
\end{abstract}

\maketitle
\section{Introduction.}

The Concordance or $\Lambda$CDM model has been successful in fitting multiple precise cosmological observations by crucially relying on a positive cosmological constant $\Lambda$, whose precise theoretical understanding remains lacking, whether it is assumed to be an approximation to a dark energy source or a useful empiric construct that fits observations  \cite{ellis2012relativistic,peebles2003cosmological}, Observational discrepancies with the $\Lambda$CDM model  \cite{aluri2023observable,raveri2019concordance} (for example, the $H_0$ tension \cite{di2021realm}) have lead to attempts to fit observations within General Relativity (GR) that avoid the need of a $\Lambda$ term, assuming large scale inhomogeneity \cite{buchert2008dark, wiltshire2008gravitational,racz2017concordance} or tilted frames  \cite{tsagas2022deceleration}. However, the shortcomings of the Concordance model are mostly addressed by the search of gravity theories alternative to GR that might lead to a more robust theoretical framework in fitting observations \cite{petrov2020introduction,clifton2012modified,PhysRevD.101.084025}, for example: $f({\cal R})$ theories \cite{de2010f,sotiriou2010f,chakraborty2022note},  torsion and teleparallel gravity \cite{shie2008torsion,bahamonde2023teleparallel}, scalar-tensors  \cite{langlois2019dark}. These theories often lead to intractable field equations and so far  exhibit shortcomings of their own \cite{PhysRevD.108.024027,koyama2016cosmological}.    

In 2021 Harada \cite{harada2021emergence} proposed a new metric gravity theory called ``Cotton Gravity'' (denoted henceforth as ``CG'') that generalizes GR through the rank-3 Cotton tensor \cite{cotton1899varietes,garcia2004cotton}. All solutions of GR are also solutions of CG, but non-trivial CG solutions can be obtained that are not GR solutions (see exchange in \cite{bargueno2021comment} and  \cite{harada2021reply}). Harada derived in  \cite{harada2021emergence} the first non-trivial CG solution: a Schwarzschild  analogue and in a second article \cite{harada2022cotton} he used the post-Newtonian limit of CG to fit galactic rotation curves without assuming the existence of dark matter. 

However,  using Harada's initial formulation does not allow to find non-vacuum conformally flat solutions, such as the FLRW models, since for these spacetimes the Cotton tensor vanishes identically \cite{garcia2004cotton}, perhaps explaining why CG has not been applied to such models.  More recently, Mantica and Molinari \cite{mantica2023codazzi} developed an alternative formulation of CG, proving that  the original field equations in \cite{harada2021emergence} are equivalent to the conditions that define as a Codazzi tensor a symmetric tensor formed with the metric and the Ricci and energy momentum tensors of GR. These authors further discussed the compatibility between CG and various assumptions on sources and their connection with known GR solutions.   

In the present letter we obtain FLRW models as non-trivial CG solutions using the formulation derived in \cite{mantica2023codazzi}, which we rewrite to make it more intuitive.  By demanding that the energy momentum tensor is not modified, we show that CG introduces as a geometric degree of freedom a time dependent term ${\cal K}$ that modifies the Ricci scalar ${}^3{\cal R}$ of the hypersurfaces orthonormal to $u^a$ (the spatial curvature), though CG theory does not provide (in general) an evolution equation for this term.

 Hence, demanding a well posed initial value formulation rules out all models, save three particular cases: (1) FLRW models of GR with $\Lambda=0$ that  follow as trivial CG solutions, (2) FLRW models equivalent to spatially flat FLRW models of GR with a constant ${}^3{\cal R}$ playing the role $\Lambda\ne 0$ and (3) as in (2), but the constant ${}^3{\cal R}$ acting as a correction to spatial curvature of GR models. The two latter cases are operationally indistinguishable from FLRW models of GR with $\Lambda\ne 0$, but the theoretical interpretation of $\Lambda$ is different: it is a constant value of spatial curvature. This covariant characterization of $\Lambda$ by CG renders the  $\Lambda$CDM model, so far a preferred reference to fit observations, as  the unique FLRW model with constant negative spatial curvature.                

\section{Cotton gravity}

The CG field equations introduced by Harada in \cite{harada2021emergence} (in geometric units $c = G = 1$) are 
\begin{equation}
    	C_{abc}=\nabla_d G^d_{abc} = 16\pi \nabla_d T^d_{abc}  ,\label{original}
\end{equation} 
where $C_{abc}$ is the Cotton tensor \cite{cotton1899varietes, stephani2009exact, garcia2004cotton} 
\begin{equation}C_{abc} = \nabla_b {\cal R}_{ac}-\nabla_c{\cal R}_{ab}-\frac16\left(g_{ac}\nabla_b{\cal R}-g_{ab}\nabla_c{\cal R}\right), \label{cotton}\end{equation}
with ${\cal R}_{a b}$ and ${\cal R}$ respectively the Ricci tensor and scalar. The connection to GR follows from  \eqref{original}, since: $g^{cd}G_{acbd} = G_{ab}$ and $g^{cd}T_{acbd} = T_{ab}$, showing that passing from GR to CG does not (necessarily) involve a change of $T^{ab}$. 

Since $C_{abc}=0$ holds for all conformally flat spacetimes (from  $C_{abc}=-4\nabla_d C_{abc}{}^d$  \cite{garcia2004cotton}), Harada proposed that condition $C_{abc}=0$ should be thought of as a CG generalization  of ${\cal R}_{ab}=0$ defining vacuum solutions in GR. However, the two conditions are distinct: ${\cal R}_{ab}=0$ is a non-trivial constraint that is often difficult to solve, while  $C_{abc}=0$ is an identity that holds automatically for all conformally flat spacetimes regardless of their sources or symmetry properties. 

However, as opposed to \eqref{original}, the alternative equivalent formulation of CG in \cite{mantica2023codazzi} does not involve explicitly computing the Cotton tensor beforehand.  It is the condition that defines a Codazzi tensor 
\begin{equation}\nabla_b{\cal C}_{ac} - \nabla_c{\cal C}_{ab} =0,\label{feqsCG}\end{equation}
for a specific simmetric tensor ${\cal C}_{ab}$ constructed with the Ricci and energy momentum tensors of GR and the metric. To make it more intuitive, we have rewritten the tensor ${\cal C}_{ab}$ in \cite{mantica2023codazzi} as
\begin{equation} {\cal C}_{ab}=G_{ab}- 8\pi T_{ab} -\frac13 (G-8\pi T)\,g_{ab}, \label{codazzi}
\end{equation}
where  $G=g^{cd}G_{cd},\,\,  T=g^{cd}T_{cd}$ and we assume that $\nabla_b T^{ab}=0$ holds.  Equation \eqref{feqsCG} makes it evident that any GR solution with $G_{ab}-8\pi T_{ab}=0$ (or ${\cal R}_{ab}=0$ if vacuum) is a CG solution, since  ${\cal C}_{ab}= 0$ holds by construction and \eqref{feqsCG} holds trivially. 

As shown in \cite{mantica2023codazzi},  a CG solution with ${\cal C}_{ab}\ne 0$ can be written as a formal solution of Einstein's equations $G^{ab} = 8\pi T^{ab}_{\hbox{\tiny{mod}}}$, with the modified energy momentum $T^{ab}_{\hbox{\tiny{mod}}}$ describing the GR source $T^{ab}$ plus an ``effective'' fluid obtained by passing to the right hand side the geometric terms brought by CG. This follows by rewriting \eqref{codazzi} as 
\begin{eqnarray}  G_{ab} = 8\pi T^{ab}_{(\hbox{\tiny{mod}})} =  8\pi\left[T^{ab}+{\cal C}^{ab}-{\cal C}\,g^{ab}\right],\quad \Rightarrow\quad \nabla_bT^{ab}_{(\hbox{\tiny{mod}})}=0, \label{TabCG}\end{eqnarray} 
where ${\cal C} = g^{cd} {\cal C}_{cd}$. This mimicking of Einstein's equations with ``effective'' fluids is common in the study of other alternative gravity theories (see examples of $f(R)$ theories in \cite{de2010f,sotiriou2010f,chakraborty2022note}). However,  these are geometric fluids not sustained by physical assumptions, hence we will not consider them. Unless specified otherwise, we will assume that all mention of matter-energy sources refer to $T^{ab}$ not to $T^{ab}_{(\hbox{\tiny{mod}})}$.

Finding non-trivial CG solutions requires that $G_{ab}-8\pi T_{ab}\ne 0$, so that ${\cal C}_{ab}\ne 0$ and  $\nabla_b T^{ab}=0$ both hold. The nonzero ${\cal C}_{ab}$ is then inserted into \eqref{feqsCG}, which yields a constraint whose solution determines a non-trivial CG solution. We illustrate this process with the Schwarzschild analogue found by Harada in  \cite{harada2021emergence}, which follows readily by applying \eqref{feqsCG}-\eqref{codazzi} to the spherically symmetric metric
\begin{equation} ds^2 = -\Phi(r) dt^2 + \frac{dr^2}{\Phi(r)} + r^2(d\theta^2+\sin^2\theta d\phi^2)\label{Schw1} \end{equation}
leading from \eqref{codazzi} to the Codazzi tensor with components   ${\cal C}^\theta_\theta={\cal C}^\phi_\phi=-2 {\cal C}^t_t =-2{\cal C}^r_r= (\Phi_{rr}r^2-2\Phi_{,r}r-4\Phi)/(6r^2)$, which inserted into  \eqref{feqsCG} yields a constraint whose solution is the same as that found by Harada in \cite{harada2021emergence}
\begin{equation}
\Phi(r) = 1 - \frac{2M_s}{r} + \gamma\,r + \frac{8\pi}{3}\,\Lambda\,r^2, \label{Schw2}
\end{equation}
with $\gamma\,r$ the extra term found by Harada and  the cosmological constant emerging as an integrating constant. We now follow the same steps to find FLRW models as CG solutions. 

\section{FLRW solutions in Cotton Gravity}

FLRW models are often described by the Robertson-Walker metric in spherical coordinates
\begin{equation} ds^2  -dt^2+a^2(t) \left[\frac{dr^2}{1-kr^2}.+r^2(d\theta^2+\sin^2\theta d\phi^2) \right], \label{flrw}\end{equation} 
where $k=k_0 H_0^2|\Omega_0^{(k)}|,\,\,k_0=0,\pm\,1$ and the source is a perfect fluid $T^{ab}=\rho u^au^b+ph^{ab},\,\,u^a=\delta^a_t$. The tensor \eqref{codazzi} for  \eqref{flrw} and $T^{ab}$  is 
\begin{equation}\fl {\cal C}^r_r = {\cal C}^\theta_\theta =  {\cal C}^\phi_\phi = -\frac{8\pi}{3}\rho +\frac{\dot a^2 +k}{a^2},\qquad 
 {\cal C}^t_t = 8\pi\left(\frac23\rho+p \right) +\frac{2a \ddot a-\dot a^2-k}{a^2}.\label{codRW} 
\end{equation}
The GR field equations $G_{ab}-8\pi T_{ab}=0$ and the balance equation $\nabla_b T^{ab}=0$ are 
\begin{eqnarray} \dot a^2 &=& \frac{8\pi}{3}\rho\,a^2-k,\label{friedeq2a}\\
\dot \rho &=& -\frac{3\dot a}{a}(\rho+p)\quad \Rightarrow\quad \frac{d\rho}{\rho+p(\rho)}=-\frac{3da}{a}, \label{friedeq2b}\end{eqnarray}
leading to, as expected, to  ${\cal C}^a_b=0$, an FLRW model of GR in which \eqref{friedeq2b} yields $\rho=\rho(a)$ allowing to obtain $a=a(t)$ by integrating \eqref{friedeq2a}.  

To obtain FLRW  models that are non-trivial CG solutions we need to assure that $G_{ab}-8\pi T_{ab}\ne 0$ and $\nabla_b T^{ab}=0$ hold for the RW metric \eqref{flrw}.  Under these conditions, the only way to modify FLRW models of GR is to modify the Friedman equation \eqref{friedeq2a}, for example as 
\begin{equation} \frac{\dot a^2}{a^2} = \frac{8\pi}{3}\rho-\frac{k+\lambda{\cal K}(t)}{a^2},\label{friedeq3}\end{equation}
where  ${\cal K}(t)$ is (for the time being) a dimensionless  free function  and $\lambda$ is a constant  with inverse squared length units to be determined further ahead. The components of the nonzero Codazzi tensor are now
\begin{equation} {\cal C}^r_r = {\cal C}^\theta_\theta =  {\cal C}^\phi_\phi = \frac{\lambda{\cal K}}{a^2},\quad {\cal C}^t_t=\frac{\lambda}{a}\left(\frac{\dot{\cal K}}{2\dot a}-\frac{{\cal K}}{a}\right),\end{equation}
which inserted in \eqref{feqsCG} yields a constraint whose solution precisely defines the new Friedman equation
\begin{equation} 8\pi\rho\,a^2-\dot a^2 -3[k+\lambda{\cal K}]= 0,\label{friedeq3b}\end{equation}
without imposing any restriction on ${\cal K}(t)$.  For any source given by $T^{ab}$ complying with the balance law \eqref{friedeq2b}, a large class of FLRW models as CG solutions follows from any choice of ${\cal K}$ in \eqref{friedeq3}. 

We emphasize that the term $\lambda {\cal K}(t)$ in \eqref{friedeq3} is not the density of another matter-energy source (as would be the case for a Friedman equation in GR). This term is a geometric degree of freedom from CG that is absent in GR.  In fact, the term ${\cal K}$ modifies the expansion scalar $\Theta=\nabla_au^a=3\dot a/a$ of the FLRW models, which in turn modifies the extrinsic curvature $K_{ab}$ of the 3-dimensional hypersurfaces of constant $t$ orthonormal to $u^a$ (which changes the way these submanifolds are embedded in the enveloping spacetime  \cite{poisson2004relativist})
\begin{equation}K_{ab} =h_a^c h_b^d \nabla_bu_a= \frac{\Theta}{3} h_{ab} =\left[\frac{8\pi}{3}\rho-\frac{k+\lambda{\cal K}}{a^2}\right]^{1/2}h_{ab}.\label{KCG1}\end{equation}
Using \eqref{KCG1} and the contracted Gauss-Codazzi equations \cite{ellis2012relativistic,poisson2004relativist,thorne2000gravitation}, we obtain the precise covariant identification of the degree of freedom brought by CG through the term ${\cal K}$: it modifies the scalar intrinsic curvature of 3-dimensional submanifolds orthonormal to $u^a$ in FLRW models:
\begin{eqnarray}\fl {}^3{\cal R}=-2\kappa T_{ab}u^au^b+ K_{ab}K^{ab}-K^2 = -16\pi\rho+\frac23\Theta^2 = \frac{6[k+\lambda{\cal K}]}{a^2}.\label{KCG2} \end{eqnarray}
where $K=K^a_a$ and   ${}^3{\cal R}$ is the Ricci scalar of the hypersurfaces orthonormal to $u^a$. 

The modification of  ${}^3{\cal R}$ produced by CG leads to solutions of the following system of 3 evolution equations and the Friedman equation \eqref{friedeq3} as the Hamiltonian constraint:
\begin{eqnarray}
\dot\rho &=& -(\rho+p)\Theta,\label{CGa}\\
\dot\Theta &=& -\frac13\Theta^2-4\pi(\rho+3p)-\frac{3\lambda}{2a}\frac{\dd {\cal K}(a)}{\dd a},\label{CGb}\\
\dot a &=& \frac13\Theta a,\label{CGc}\\
\frac{\Theta^2}{9} &=& \frac{8\pi}{3}\rho-\frac{k+\lambda {\cal K}}{a^2},\label{CGd}
\end{eqnarray} 
whose integration requires the specification of ${\cal K}={\cal K}(a)$. Since there is no evolution equation or conservation law for this term, then for a general ${\cal K}(a)$ the system \eqref{CGa}-\eqref{CGd} is incompatible with a well posed initial value formulation that requires finding a unique solution given initial data at an initial hypersurface $t=t_0$ for which $a(t_0)=1$. In other words, in general, initial data at $t=t_0$ determines only ${\cal K}_0={\cal K}(a(t_0))={\cal K}(1)$, which without an evolution equation does not determine ${\cal K}$ for $t\ne t_0$. While ${\cal K}(a)$ can be prescribed beforehand and could lead to interesting dynamics unachievable by FLRW models in GR, this involves a priori determination of future (even asymptotic) evolution disconnected from initial conditions.

However, there are three particular cases of FLRW models in CG in which \eqref{CGa}-\eqref{CGd} are compatible with a well posed initial value formulation:
\begin{enumerate}
\item ${\cal K}=$ constant. These are the FLRW models of GR with $\Lambda=0$ (the constant can be absorbed into a redefinition of $k$). If  ${\cal K}=0$ these are trivial solutions of CG with ${\cal C}^a_b=0$. 
\begin{equation}H^2 =\frac{\dot a^2}{a^2}=\frac{8\pi}{3}\rho-\frac{k}{a^2},\label{mod1}\end{equation}
Since ${}^3{\cal R}=6k/a^2$, the only case with constant (zero) spatial curvature is $k=0$.
\item $k=0$ and ${\cal K}=a^2$. This choice leads to constant spatial curvature: ${}^3{\cal R}=6\lambda$, whose sign is determined by the sign of $\lambda$. These are non-trivial CG solutions with  ${\cal C}^a_b\ne 0$. 
\begin{equation}H^2 =\frac{\dot a^2}{a^2}=\frac{8\pi}{3}\rho-\lambda,\label{mod2}\end{equation}
The models coincide with models that are spatially flat in GR with $\Lambda\ne 0$ by identifying $\lambda=(8\pi/3)\Lambda$.
\item $k\ne 0$ and ${\cal K}=a^2$. It leads to FLRW models in which $\lambda$ plays the role of a cosmological constant, but it is a constant correction to the varying spatial curvature ${}^3{\cal R}=6k/a^2$ of GR models. Since the sign of the spatial curvature determines the standard topology of the hypersurfaces orthonormal to $u^a$, we assume that $k$ and $\lambda$ have same signs. 
\begin{equation}H^2 =\frac{\dot a^2}{a^2}=\frac{8\pi}{3}\rho-\frac{k}{a^2}-\lambda,\label{mod3}\end{equation}
The models coincide with FLRW models with $\Lambda\ne 0$ identifying  $\lambda=(8\pi/3)\Lambda$. In the case of negative spatial curvature  $\lambda<0$ is an asymptotic residual spatial curvature as $a\to\infty$.
 \end{enumerate}
While all FLRW models \eqref{mod1}-\eqref{mod3} operationally coincide with FLRW models of GR, there is an important conceptual difference. In GR, $\Lambda$ has units of energy density, so theoretically it is a source (likely an empiric approximation of dark energy), hence it is normalized by $8\pi G/c^4$ to put it in 1/$\hbox{length}^2$ units. In CG $\lambda$ has the same dynamical effect as $(8\pi/3)\Lambda$, but it is a constant spatial curvature, not a source.  Hence, to emphasize the conceptual difference it is better to use a different symbol $\lambda$ and since it already has  1/$\hbox{length}^2$ units, it can be fixed simply as a value proportional to $H_0^2$ that provides a best fit for observations.  

Since  $\Lambda>0$ is assumed to hold in cosmology and $\rho$ can be assumed to represent cold dark matter (CDM) density (neglecting other constituents like baryons, electrons, photons and neutrinos),  then the cases of interest are \eqref{mod2} and \eqref{mod3} with $\lambda<0$ and assigning the numerical value $\lambda = -(8\pi/3)\Lambda$ determined by observations. We have for these cases the Hubble scalar in terms of CDM and spatial curvature without dark energy or a $\Lambda$ in energy density units: 
\begin{eqnarray}\fl  \frac{H^2}{H_0^2} &=& \Omega^{(\hbox{\tiny{CDM}})}+ \Omega^{(K)},\qquad   \Omega^{(\hbox{\tiny{CDM}})}=\frac{\Omega_0^{(\hbox{\tiny{CDM}})}}{a^3},\quad \Omega^{(K)}=\frac{\Omega_0^{(k)}}{a^2}+\Omega_0^{(\lambda)},\label{CGH1}\\ 
\fl \Omega_0^{(m)} &=&\frac{8\pi\rho_0}{3H_0^2},\quad \Omega_0^{(k)} =\frac{|k|}{H_0^2},\quad \Omega_0^{(\lambda)} =\frac{|\lambda|}{H_0^2}\label{CGH2}\end{eqnarray}
with $\Omega_0^{(K)}=\Omega_0^{(k)}+\Omega_0^{(\lambda)}=1-\Omega_0^{(\hbox{\tiny{CDM}})}$. In particular,  the $\Lambda$CDM model is equivalent to the FLRW model in  \eqref{mod2} or \eqref{CGH1} with $\Omega_0^{(k)} =0$, which is the unique FLRW dust model in CG with constant negative spatial curvature. This theoretical characterization of the $\Lambda$CDM model is more appealing than that of GR, either as a spatially flat  FLRW model providing a late time description of an accelerated cosmic expansion dominated by a dark energy source approximated by $\Lambda$, or simply as an FLRW model empirically constructed to fit observations. In fact, the FLRW models  \eqref{mod2} and \eqref{mod3} can fit observations and account for an accelerated expansion driven by spatial curvature, without the need to invoke a dark energy source. We can also identify in  \eqref{mod2} the constant curvature vacuum cases setting $\rho=0$, lading to de Sitter ($\lambda = -(8\pi/3)\Lambda$), anti-de Sitter ($\lambda = (8\pi/3)\Lambda$) and Minkwski ($\lambda=0$). 

The ability of non-trivial FLRW models of CG in accounting for late time cosmic acceleration and fitting observations through spatial curvature can only be reproduced in the Concordance model of GR by the loosely understood cosmological constant, or by dynamic dark energy models compatible with negative pressure (quintessence, Chaplygin gas or fluid models).  Since the CG models we have presented are operationally indistinguishable from FLRW models of GR, practically all theoretical and observational computations in FLRW models of GR can be directly applied, just bearing in mind that $\Lambda$ is now spatial curvature, However, much more work is required, for example cosmological perturbations and structure formation,  to understand the impact of CG on cosmic dynamics. So we do not claim that our results prove that CG is a preferred theory over GR or that they allow to eliminate the need of assuming a dark energy source. Rather, our aim is to illustrate the potential of CG to provide a valuable alternative to GR, and valuable theoretical arguments to better understand the dark sector, as well as to probe alternative dynamical approaches \cite{buchert2008curvature,desgrange2019dynamical,heinesen2020solving}. \\


\section*{Acknowledgments}
SN acknowledges financial support from SEP–CONACYT postgraduate grants program.\\

\section*{References}


\bibliographystyle{unsrt}
\bibliography{CottonFLRWLetter}       

\begin{thebibliography}{10}

\bibitem{ellis2012relativistic}
George~FR Ellis, Roy Maartens, and Malcolm~AH MacCallum.
\newblock {\em Relativistic cosmology}.
\newblock Cambridge University Press, 2012.

\bibitem{peebles2003cosmological}
P~James~E Peebles and Bharat Ratra.
\newblock The cosmological constant and dark energy.
\newblock {\em Reviews of modern physics}, 75(2):559, 2003.

\bibitem{aluri2023observable}
Pavan~Kumar Aluri, Paolo Cea, Pravabati Chingangbam, Ming-Chung Chu, Roger~G
  Clowes, Damien Hutsem{\'e}kers, Joby~P Kochappan, Alexia~M Lopez, Lang Liu,
  Niels~CM Martens, et~al.
\newblock Is the observable universe consistent with the cosmological
  principle?
\newblock {\em Classical and Quantum Gravity}, 40(9):094001, 2023.

\bibitem{raveri2019concordance}
Marco Raveri and Wayne Hu.
\newblock Concordance and discordance in cosmology.
\newblock {\em Physical Review D}, 99(4):043506, 2019.

\bibitem{di2021realm}
Eleonora Di~Valentino, Olga Mena, Supriya Pan, Luca Visinelli, Weiqiang Yang,
  Alessandro Melchiorri, David~F Mota, Adam~G Riess, and Joseph Silk.
\newblock In the realm of the hubble tension—a review of solutions.
\newblock {\em Classical and Quantum Gravity}, 38(15):153001, 2021.

\bibitem{buchert2008dark}
Thomas Buchert.
\newblock Dark energy from structure: a status report.
\newblock {\em General Relativity and Gravitation}, 40:467--527, 2008.

\bibitem{wiltshire2008gravitational}
David~L Wiltshire.
\newblock Gravitational energy and cosmic acceleration.
\newblock {\em International Journal of Modern Physics D}, 17(03n04):641--649,
  2008.

\bibitem{racz2017concordance}
G{\'a}bor R{\'a}cz, L{\'a}szl{\'o} Dobos, R{\'o}bert Beck, Istv{\'a}n Szapudi,
  and Istv{\'a}n Csabai.
\newblock Concordance cosmology without dark energy.
\newblock {\em Monthly Notices of the Royal Astronomical Society: Letters},
  469(1):L1--L5, 2017.

\bibitem{tsagas2022deceleration}
Christos~G Tsagas.
\newblock The deceleration parameter in ‘tilted’universes: generalising the
  friedmann background.
\newblock {\em The European Physical Journal C}, 82(6):521, 2022.

\bibitem{petrov2020introduction}
Albert Petrov.
\newblock {\em Introduction to modified gravity}.
\newblock Springer Nature, 2020.

\bibitem{clifton2012modified}
Timothy Clifton, Pedro~G Ferreira, Antonio Padilla, and Constantinos Skordis.
\newblock Modified gravity and cosmology.
\newblock {\em Physics reports}, 513(1-3):1--189, 2012.

\bibitem{PhysRevD.101.084025}
Nils Alex and Tobias Reinhart.
\newblock Covariant constructive gravity: A step-by-step guide towards
  alternative theories of gravity.
\newblock {\em Phys. Rev. D}, 101:084025, Apr 2020.

\bibitem{de2010f}
Antonio De~Felice and Shinji Tsujikawa.
\newblock f (r) theories.
\newblock {\em Living Reviews in Relativity}, 13(1):1--161, 2010.

\bibitem{sotiriou2010f}
Thomas~P Sotiriou and Valerio Faraoni.
\newblock f (r) theories of gravity.
\newblock {\em Reviews of Modern Physics}, 82(1):451, 2010.

\bibitem{chakraborty2022note}
Saikat Chakraborty, Peter~KS Dunsby, and Kelly Macdevette.
\newblock A note on the dynamical system formulations in f (r) gravity.
\newblock {\em International Journal of Geometric Methods in Modern Physics},
  19(08):2230003, 2022.

\bibitem{shie2008torsion}
Kun-Feng Shie, James~M Nester, and Hwei-Jang Yo.
\newblock Torsion cosmology and the accelerating universe.
\newblock {\em Physical Review D}, 78(2):023522, 2008.

\bibitem{bahamonde2023teleparallel}
Sebastian Bahamonde, Konstantinos~F Dialektopoulos, Celia Escamilla-Rivera,
  Gabriel Farrugia, Viktor Gakis, Martin Hendry, Manuel Hohmann, Jackson~Levi
  Said, Jurgen Mifsud, and Eleonora Di~Valentino.
\newblock Teleparallel gravity: from theory to cosmology.
\newblock {\em Reports on Progress in Physics}, 86(2):026901, 2023.

\bibitem{langlois2019dark}
David Langlois.
\newblock Dark energy and modified gravity in degenerate higher-order
  scalar--tensor (dhost) theories: A review.
\newblock {\em International Journal of Modern Physics D}, 28(05):1942006,
  2019.

\bibitem{PhysRevD.108.024027}
Qing Gao, Yujie You, Yungui Gong, Chao Zhang, and Chunyu Zhang.
\newblock Testing alternative theories of gravity with space-based
  gravitational wave detectors.
\newblock {\em Phys. Rev. D}, 108:024027, Jul 2023.

\bibitem{koyama2016cosmological}
Kazuya Koyama.
\newblock Cosmological tests of modified gravity.
\newblock {\em Reports on Progress in Physics}, 79(4):046902, 2016.

\bibitem{harada2021emergence}
Junpei Harada.
\newblock Emergence of the cotton tensor for describing gravity.
\newblock {\em Physical Review D}, 103(12):L121502, 2021.

\bibitem{cotton1899varietes}
{\'E}mile Cotton.
\newblock Sur les vari{\'e}t{\'e}s a trois dimensions.
\newblock In {\em Annales de la Facult{\'e} des sciences de Toulouse:
  Math{\'e}matiques}, volume~1, pages 385--438, 1899.

\bibitem{garcia2004cotton}
Alberto~A Garc{\'\i}a, Friedrich~W Hehl, Christian Heinicke, and Alfredo
  Macias.
\newblock The cotton tensor in riemannian spacetimes.
\newblock {\em Classical and Quantum Gravity}, 21(4):1099, 2004.

\bibitem{bargueno2021comment}
Pedro Bargue{\~n}o.
\newblock Comment on “emergence of the cotton tensor for describing
  gravity”.
\newblock {\em Physical Review D}, 104(8):088501, 2021.

\bibitem{harada2021reply}
Junpei Harada.
\newblock Reply to “comment on ‘emergence of the cotton tensor for
  describing gravity”’.
\newblock {\em Physical Review D}, 104(8):088502, 2021.

\bibitem{harada2022cotton}
Junpei Harada.
\newblock Cotton gravity and 84 galaxy rotation curves.
\newblock {\em Physical Review D}, 106(6):064044, 2022.

\bibitem{mantica2023codazzi}
Carlo~Alberto Mantica and Luca~Guido Molinari.
\newblock Codazzi tensors and their space-times and cotton gravity.
\newblock {\em General Relativity and Gravitation}, 55(4):62, 2023.

\bibitem{stephani2009exact}
Hans Stephani, Dietrich Kramer, Malcolm MacCallum, Cornelius Hoenselaers, and
  Eduard Herlt.
\newblock {\em Exact solutions of Einstein's field equations}.
\newblock Cambridge university press, 2009.

\bibitem{poisson2004relativist}
Eric Poisson.
\newblock {\em A relativist's toolkit: the mathematics of black-hole
  mechanics}.
\newblock Cambridge university press, 2004.

\bibitem{thorne2000gravitation}
Kip~S Thorne, John~Archibald Wheeler, and Charles~W Misner.
\newblock {\em Gravitation}.
\newblock Freeman San Francisco, CA, 2000.

\bibitem{buchert2008curvature}
Thomas Buchert and Mauro Carfora.
\newblock On the curvature of the present-day universe.
\newblock {\em Classical and Quantum Gravity}, 25(19):195001, 2008.

\bibitem{desgrange2019dynamical}
C{\'e}lia Desgrange, Asta Heinesen, and Thomas Buchert.
\newblock Dynamical spatial curvature as a fit to type ia supernovae.
\newblock {\em International Journal of Modern Physics D}, 28(11):1950143,
  2019.

\bibitem{heinesen2020solving}
Asta Heinesen and Thomas Buchert.
\newblock Solving the curvature and hubble parameter inconsistencies through
  structure formation-induced curvature.
\newblock {\em Classical and Quantum Gravity}, 37(16):164001, 2020.

\end{thebibliography}

\end{document}